\documentclass[10pt,conference]{IEEEtran}
\IEEEoverridecommandlockouts
\usepackage{cite}
\usepackage{amsmath,amssymb,amsfonts}
\usepackage{algorithmic}
\usepackage{graphicx}
\usepackage{textcomp}
\usepackage{xcolor}
\usepackage{enumerate}
\usepackage{array,booktabs,arydshln,xcolor}
\usepackage[hidelinks]{hyperref}
\usepackage{caption}
\usepackage{comment}
\usepackage{siunitx}
\usepackage{stfloats}
\usepackage{xurl}

\newcommand{\cmmnt}[1]{}

\def\BibTeX{{\rm B\kern-.05em{\sc i\kern-.025em b}\kern-.08em
    T\kern-.1667em\lower.7ex\hbox{E}\kern-.125emX}}
    
\begin{document}

\title{Leveraging Interpretable Tsetlin Machine for \\ PDF Malware Detection \\
}

\author{
    \IEEEauthorblockN{Rahul Jaiswal, Ole-Christoffer Granmo\\
    The Centre for Artificial Intelligence Research (CAIR) \\
    \IEEEauthorblockA{Department of ICT, University of Agder, Norway}
{\{rahul.jaiswal, ole.granmo\}@uia.no}
}}

\IEEEoverridecommandlockouts \IEEEpubid{\makebox[\columnwidth]{979-8-3315-1276-8/26/\$31.00 \copyright 2026 IEEE \hfill} \hspace{\columnsep}\makebox[\columnwidth]{ }}

\maketitle

\begin{abstract} 
In the digital era, Portable Document Format (PDF) is one of the most widely used file formats for storing and exchanging digital documents due to its platform independence and rich functionality. However, these same capabilities have also made PDF files an attractive attack vector for cyberattackers, who embed malicious code within seemingly legitimate documents to compromise target systems. This paper presents a novel interpretable Tsetlin Machine (TM)-based framework for PDF malware detection. The proposed framework extracts salient features from PDF documents through static analysis without executing the files and employs rule-based learning to accurately classify benign and malicious PDF documents. Numerical evaluation on the RIT-PDFMal-2026 dataset demonstrates that the proposed framework achieves an accuracy of 98.02\%, outperforming several state-of-the-art machine learning classifiers. Moreover, the proposed framework provides intrinsic interpretability by transparently explaining its classification decisions. Edge deployment on a Raspberry Pi further supports real-time, on-device PDF malware detection. The combination of better accuracy, computational efficiency, and intrinsic interpretability makes the proposed framework a promising solution for practical PDF malware detection.
\end{abstract}


\section{Introduction} 
\label{intro}
\begin{figure}[t!]
\centering
\includegraphics[width=\columnwidth,height=5.0cm,keepaspectratio]{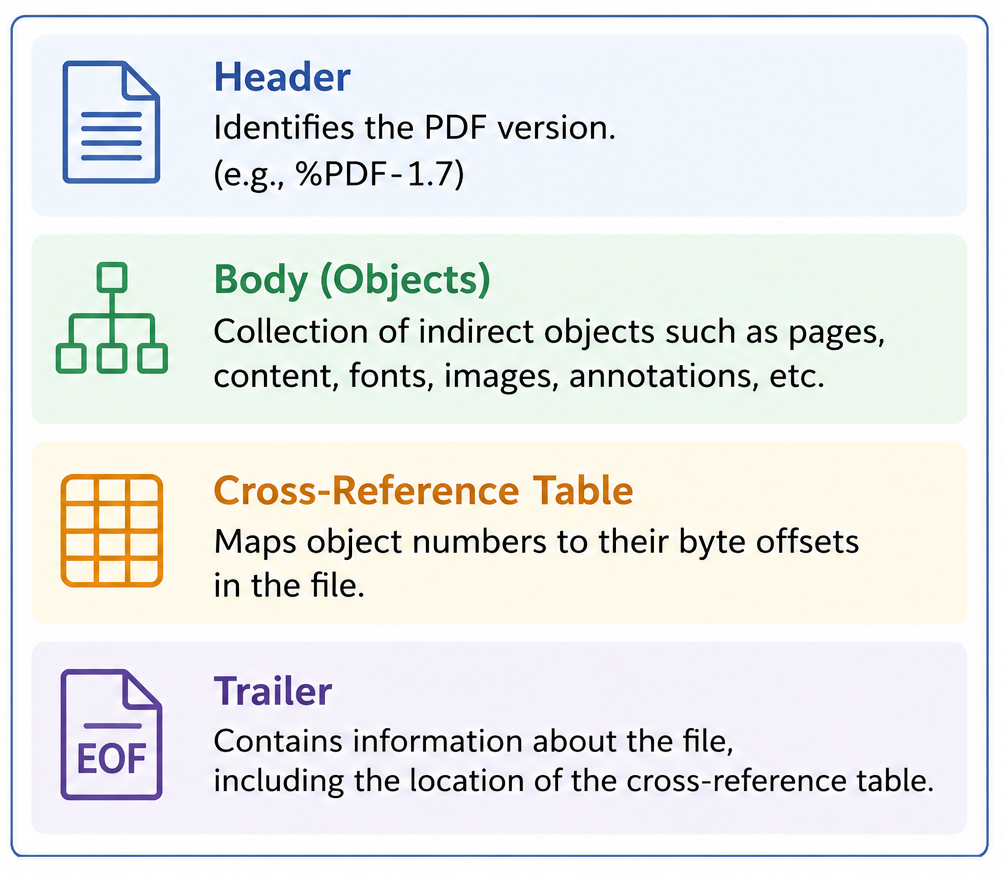} 
\caption{The PDF internal architecture.}
\label{pdf_arch} \vspace{-3mm}
\end{figure}

In today's digital world, the Portable Document Format (PDF) has become one of the most widely used document formats for sharing and exchanging information due to its portability, platform independence, and consistent rendering across different operating systems and software environments. The PDF files contain a complex internal structure consisting of both binary and ASCII elements and support advanced features such as embedded objects, JavaScript, and interactive actions, as shown in Fig.~\ref{pdf_arch}. Consequently, they can execute complex instructions when opened, extending their functionality beyond that of conventional static documents.

The CloudFiles Report 2025~\cite{report_cloudfiles} states that approximately 15 trillion digital files were generated worldwide across various formats, including PDF, doc, images, videos, and graphic designs. Among these, PDF documents account for nearly 2.5 trillion files, representing around 17\% of the total. The PDF files are widely used to store and share various types of documents, such as invoices, payslips, certificates, contracts, and reports. This widespread adoption across both personal and organizational applications has made PDF files one of the most prevalent formats for digital document exchange.

The widespread adoption of PDF documents and their advanced functionalities have made them an attractive target for cyberattackers. Features such as embedded objects and JavaScript can be exploited to deliver malicious payloads, making PDF files a common attack vector for malware distribution. Malicious PDFs can facilitate cyberattacks such as credential theft, spyware installation, unauthorized system access, browser exploitation, data exfiltration, phishing, and financial fraud~\cite{singh2020malware}. Moreover, the rapid evolution of different attack techniques makes PDF malware detection a significant challenge for modern cybersecurity systems. The Reis Informatica Report 2026~\cite{report_reis} highlights that 74\% of cyberattacks against Microsoft Windows systems in Canada were carried out via malicious PDF documents.

To protect PDF documents, a variety of malware detection techniques are used. For example, signature-based methods~\cite{singh2020malware} identify malware by matching files against known signatures, such as code patterns, hashes, or predefined behavioral characteristics. However, they struggle to identify newly emerging malware. Anomaly-based methods~\cite{liu2025vapd} learn the characteristics of normal files or system behavior and detect deviations. By relying on behavioral anomalies, these methods can identify unknown and evolving malware. Recently, machine learning (ML) techniques, such as decision trees, random forest, support vector machine, and gradient boosting, have been explored for PDF malware detection~\cite{abu2022pdf, chbib2024leveraging}. Transfer learning~\cite{jaiswal2025leveraging, jaiswal2023location, jaiswal2025data} has also been used for detecting malware~\cite{rong2020transnet}. Some of these studies have employed post-hoc explainability techniques, such as Shapley Additive Explanations (SHAP) and Local Interpretable Model-agnostic Explanations (LIME)~\cite{lundberg2017unified}, to interpret the predictions of black-box ML classifiers. However, these techniques provide only approximate explanations rather than revealing the classifiers' underlying decision-making process. 

\begin{figure*}[b!]
\centering
\includegraphics[width=0.9\textwidth]{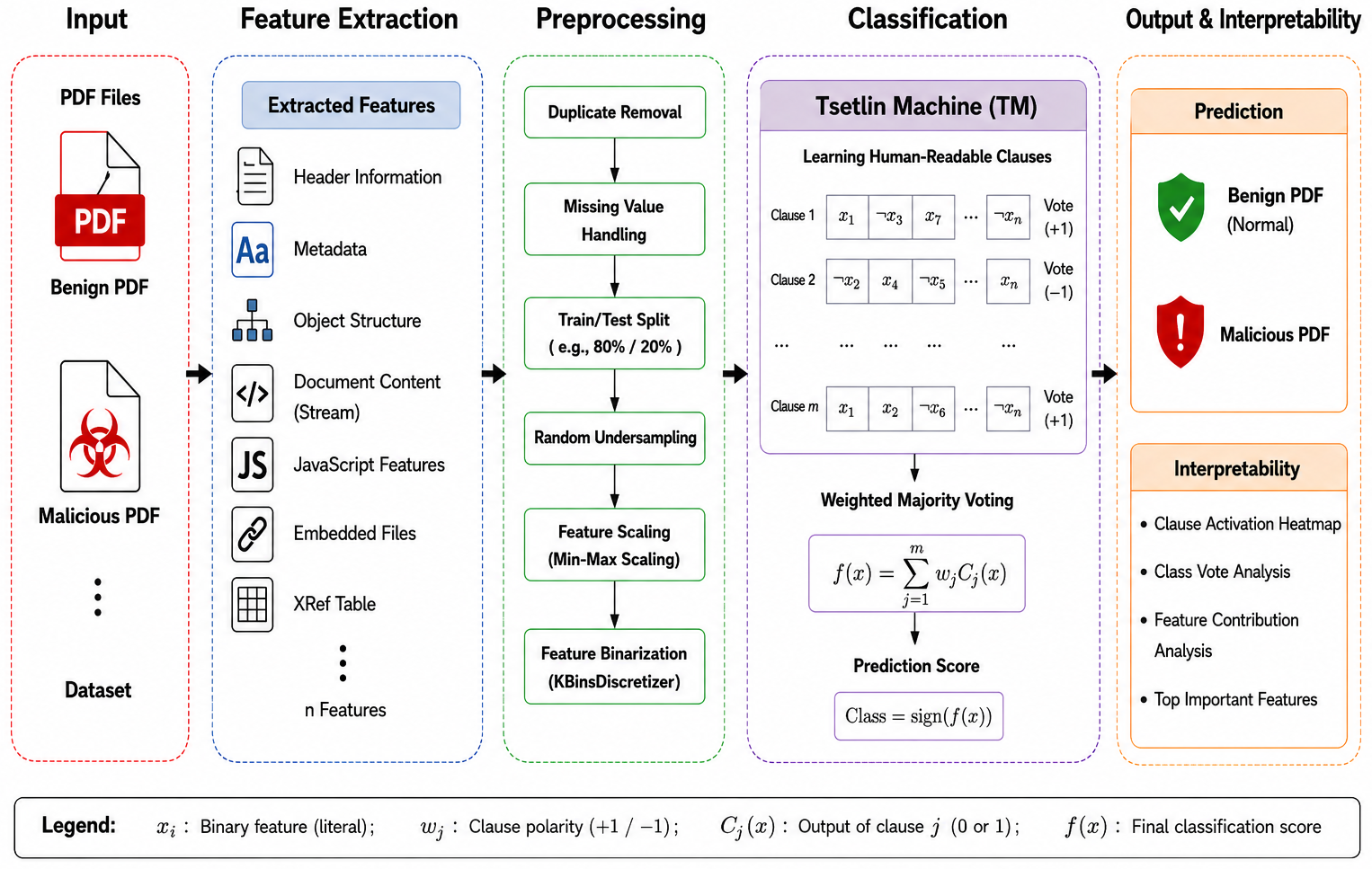}
\caption{Proposed TM framework for PDF malware detection.}
\label{tm_framwork} 
\end{figure*}

This paper proposes a novel approach for PDF malware detection using an interpretable ML model based on the Tsetlin Machine (TM)~\cite{kundu2026comprehensive, granmo2018tsetlin}. The proposed framework extracts salient features directly from PDF files without executing them and uses these features to classify documents as \texttt{benign (normal)} or \texttt{malicious}. Furthermore, unlike conventional black-box ML classifiers, the framework provides intrinsic interpretability by learning human-readable propositional clauses. Its classification decisions are directly explained through clause activation heatmaps, class-vote analysis, and feature-level contribution analysis, providing transparent and faithful insights that enhance the trustworthiness of PDF malware detection. Moreover, the framework is deployed on a Raspberry Pi edge device to demonstrate its capability for real-time, on-device inference for PDF malware detection. The key contributions of this paper are:
\begin{itemize}
    \item Design of an effective TM framework for PDF malware detection using the RIT-PDFMal-2026 dataset.
    \item Numerical evaluation showing better performance of the proposed TM framework over existing ML classifiers.
    \item Interpretability analysis of the proposed TM framework through learned clauses and feature contributions, explaining its decision-making for PDF malware detection.
    \item Deployment of the proposed TM framework on a Raspberry Pi for real-time, on-device PDF malware detection.
\end{itemize}

The rest of this paper is structured as follows. Section~\ref{bg} describes different classifiers employed. Section~\ref{prop_fram} describes the proposed TM framework. Section~\ref{exp_data} introduces the experimental dataset. Section~\ref{res_dis} presents and discusses the results. Finally, Section~\ref{con_fut} concludes the paper with future work.

\section{Background} 
\label{bg}
This section describes the Tsetlin Machine, machine learning classifiers, and edge-deployment method used in this study.

\subsection{Tsetlin Machine}
\label{tmc}
The Tsetlin Machine (TM) is an interpretable, rule-based machine learning model that learns human-readable logical clauses using propositional logic~\cite{granmo2018tsetlin}. By representing malicious cyberattack patterns as logical expressions, the TM enables transparent and explainable malware detection. The TM represents knowledge using a collection of conjunctive clauses (see Fig.~\ref{tm_framwork}) formed from binary input features. Each clause $C_j$ consists of a conjunction of selected literals and their negations, and is defined as~\cite{granmo2018tsetlin}:
\begin{equation}
C_j = \bigwedge_{k \in I_j} x_k \ \wedge \ \bigwedge_{l \in \bar{I}_j} \neg x_l,
\label{eq:clause}
\end{equation}
where $x_k \in \{0,1\}$ denotes a binary feature, while $I_j$ and $\bar{I}_j$ represent the sets of included and negated literals, respectively.

Each clause contributes either a positive or a negative vote toward a class, and the final class score is obtained by aggregating the votes from all clauses as:
\begin{equation}
f(\mathbf{x})=\sum_{j=1}^{m} w_j C_j(\mathbf{x}), \qquad |f(\mathbf{x})|\leq T,
\label{eq:voting}
\end{equation}
where $w_j \in \{+1,-1\}$ denotes the polarity of clause $j$, and $T$ is the voting threshold that constrains the accumulated clause votes to promote stable learning. The specificity parameter $s>1$ controls the granularity of the learned clauses by regulating the probability of including literals during training.

\subsection{Machine Learning Classifiers} 
\label{mlc}
In this study, six ML classifiers are considered for performance comparison with the proposed TM framework. These classifiers include Decision Tree (DT), which constructs a hierarchical tree by recursively splitting the feature space based on decision rules~\cite{breiman2017classification}; K-Nearest Neighbours (KNN), which classifies a sample according to the labels of its nearest neighbours in the feature space~\cite{alpaydin2020introduction}; Naive Bayes (NB), a probabilistic classifier derived from Bayes' theorem under the assumption of conditional feature independence~\cite{alpaydin2020introduction}, Logistic Regression (LR), which models class membership probabilities using a logistic function~\cite{hosmer2013applied}; XGBoost, a gradient-boosting algorithm that incrementally builds decision trees to improve predictive accuracy~\cite{chen2016xgboost}; and LightGBM (LGBM), a histogram-based gradient-boosting framework that adopts a leaf-wise tree growth strategy to achieve efficient training and high predictive performance~\cite{ke2017lightgbm}.

\subsection{Edge Deployment} 
\label{edm}
\begin{figure}[t!]
\centering
\includegraphics[width=\columnwidth,height=5.0cm,keepaspectratio]{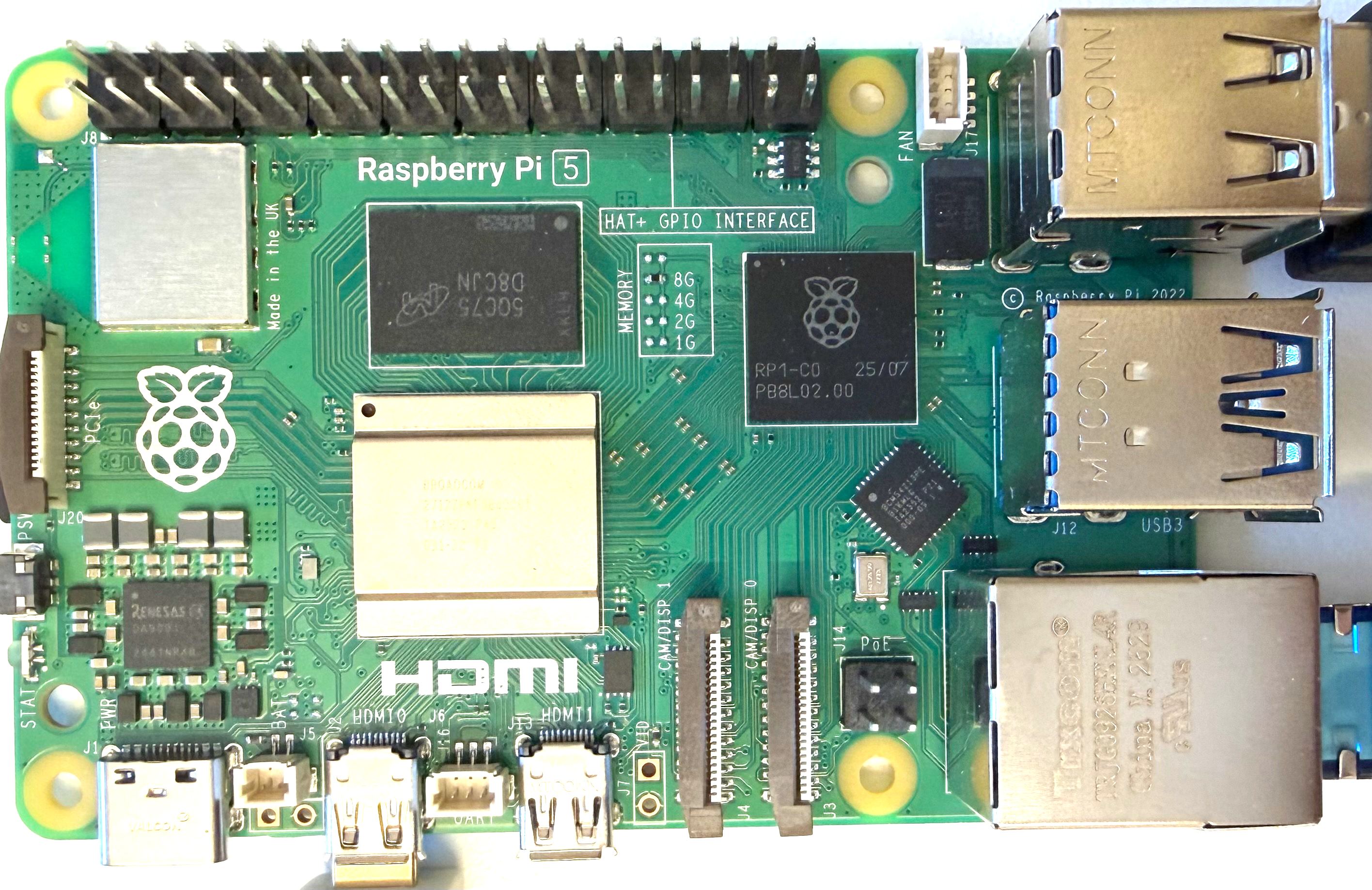} 
\caption{Experimental setup using Raspberry Pi 5 Model B.}
\label{rasp_fig} \vspace{-3mm}
\end{figure}

To validate the practical deployment of the proposed TM framework, Raspberry Pi 5 Model B~\cite{mathe2024comprehensive} is used as the target edge platform. Approximately the size of a credit card, the Raspberry Pi 5 integrates a 2.4~GHz quad-core ARM Cortex-A76 processor, up to 8~GB of SDRAM, USB 3.0 and USB 2.0 interfaces, Gigabit Ethernet, dual-band Wi-Fi, and Bluetooth connectivity (see Fig.~\ref{rasp_fig}). The trained TM model is executed directly on this device to perform real-time, on-device PDF malware detection, demonstrating that accurate inference can be achieved on a resource-constrained platform without relying on cloud-based computation.

To support edge deployment, the Raspberry Pi is connected to the host computer using an Ethernet interface, ensuring reliable communication between the two devices. The TM and ML models are implemented in Python on the Raspberry Pi to perform real-time, on-device PDF malware detection.

\section{Proposed TM Framework} 
\label{prop_fram}
The proposed TM framework aims to distinguish malicious PDF documents from benign ones accurately. As illustrated in Fig.~\ref{tm_framwork}, the framework consists of several sequential stages, including feature extraction, preprocessing, classification, and interpretability analysis.

In the first stage, salient features are extracted directly from PDF documents through static analysis without executing the files. The extracted features are then preprocessed by removing duplicate samples, handling missing values, performing data splitting, addressing class imbalance through random undersampling~\cite{mishra2017handling}, applying feature normalization, and converting the features into a binary representation using feature binarization. The processed data are subsequently used to train the TM model, where appropriate hyperparameters are selected, and logical clauses are learned to capture discriminative patterns for malware detection. To improve the robustness and generalization of the model, $k$-fold cross-validation~\cite{bhagwat2019applied} is employed during training. Finally, the trained TM model classifies an unseen PDF document as either \texttt{benign} or \texttt{malicious} by aggregating the votes of the learned positive and negative clauses. 

Next, following the training, the TM model is serialized together with its learned clauses, automata states, and preprocessing components for edge deployment. On the Raspberry Pi, the pre-trained TM model, feature binarizer, scaler, and test samples are loaded to reproduce the inference pipeline used during training. The incoming PDF feature vectors are preprocessed and classified directly as either \texttt{benign} or \texttt{malicious}. Executing the inference process locally on the edge device enables low-latency PDF malware detection while minimizing dependence on centralized computing resources.

\section{Experimental Dataset} 
\label{exp_data}
For PDF malware detection, a dataset comprising both benign and malicious PDF files is required. The RIT-PDFMal-2026 dataset\footnote{Dataset link: https://github.com/Mo-Alani/RIT-PDFMal-2026 (accessed on July 05, 2026).}~\cite{alani2026rit} is used to detect PDF malware. The dataset contains real-world malicious PDF samples collected between 2017 and 2025 from VirusTotal~\cite{link_vt}. The benign PDF files were gathered using a dedicated internet crawler that automatically downloaded PDF documents from different websites. The PDF file sizes range from 25~kB to 1.5~MB. The dataset is imbalanced and comprises 24,337 PDF samples described by 42 extracted numerical features, as presented in Table~\ref{summary_feature} and Table~\ref{summary_sample}, respectively.

\begin{table*} [t!]
\caption{Features present in the RIT-PDFMal-2026 dataset.}
\label{summary_feature} 
\centering
{
\setlength\tabcolsep{3.0pt}
\begin{tabular}{|c|c|c|c|c|c|c|c|c|c|}
\hline
S.No. & Feature name & S.No. & Feature name & S.No. & Feature name & S.No. & Feature name & S.No. & Feature name \\
\hline
1. & \texttt{pdfsize} & 2. & \texttt{metadata size} & 3. & \texttt{pages} & 4. & \texttt{xref length} & 5. & \texttt{title length} \\ 
6. & \texttt{isEncrypted} & 7. & \texttt{embedded files} & 8. & \texttt{images} & 9. & \texttt{contains\_text} & 10. & \texttt{pdf\_ver} \\  
11. & \texttt{obj} & 12. & \texttt{endobj} & 13. & \texttt{stream} & 14. & \texttt{endstream} & 15. & \texttt{trailer} \\ 
16. & \texttt{xref} & 17. & \texttt{startxref} & 18. & \texttt{page\_command} & 19. & \texttt{Encrypt} & 20. & \texttt{ObjStm} \\
21. & \texttt{JS} & 22. & \texttt{JavaScript} & 23. & \texttt{AA} & 24. & \texttt{OpenAction} & 
25. & \texttt{Acroform} \\ 
26. & \texttt{JBIG2Decode} & 27. & \texttt{RichMedia} & 28. & \texttt{Launch} & 29. & \texttt{EmbeddedFile} & 30. & \texttt{XFA} \\
31. & \texttt{Colors} & 32. & \texttt{URI} & 33. & \texttt{BaseEncoding} & 34. & \texttt{Encoding} & 35. & \texttt{ProcSet} \\ 
36. & \texttt{Registry} & 37. & \texttt{Resources} & 38. & \texttt{www} & 39. & \texttt{server} & 40. & \texttt{Root} \\ 
\cline{5-10}
41. & \texttt{BitsPerComponent} & 42. & \texttt{Label} & \multicolumn{6}{c|}{Total number of features = 42}  \\
\hline
\end{tabular} \vspace{-3mm}
}
\end{table*}

\begin{table} [t!]
\caption{Samples in the RIT-PDFMal-2026 dataset.}
\label{summary_sample} 
\centering
{
\setlength\tabcolsep{4.0pt}
\begin{tabular}{|l|c|c|c|}
\hline
Samples & Collected & Corrupted & Final \\
\hline
Benign & 13,242 & 7 & 13,235 \\
\hline
Malicious & 11,243 & 141 & 11,102 \\
\hline
Total samples & \multicolumn{3}{c|}{13,235 + 11,102 = 24,337} \\
\hline
\end{tabular} \vspace{-3mm}
}
\end{table}

\section{Results and Discussions} 
\label{res_dis}
This section outlines the experimental setup, performance evaluation and discusses the classification results.

\subsection{Experimental Setup} 
\label{sys_set}
All algorithms are implemented in Python~3.13.6. The ML models are developed using Keras built on TensorFlow~2.20.0, while NumPy~2.3.2, Pandas~2.3.1, scikit-learn~1.7.2, imbalanced-learn~0.14.0, XGBoost~3.2.0, and LightGBM~4.6.0 are used for data preprocessing and performance evaluation. All experiments are conducted on a MacBook powered by an Apple M4 chip with 16~GB of RAM. The edge deployment experiments are performed on a Raspberry Pi~5 running Python~3.11.9 and TMU~0.6.5.

\subsection{Performance Evaluation} 
\label{eva_method} 
The classification performance is evaluated using accuracy, macro-averaged precision, recall, and F1-score, which assign equal importance to each class and are therefore well suited for imbalanced datasets. Accuracy quantifies the overall proportion of correctly classified samples. Precision measures the proportion of correctly identified malicious PDF files, recall evaluates the ability to detect actual malicious PDF files, and the F1-score provides a balanced assessment by combining precision and recall into a single metric. In addition, class-wise precision, recall, and F1-score are reported for the proposed TM framework. The dataset is divided into 80\% training and 20\% testing subsets using stratified random sampling with a fixed random seed of 42. The training data are subsequently balanced through random undersampling, normalized, and binarized before applying five-fold stratified cross-validation~\cite{bhagwat2019applied} for hyperparameter selection. The final TM model is then trained on the complete preprocessed training set and evaluated on the independent test set. Furthermore, a confusion matrix is used to examine class-wise prediction performance. The proposed TM framework is benchmarked against the ML classifiers described in Section~\ref{mlc}. Finally, the inference time, defined as the average time required to classify a single input sample, is measured to evaluate computational efficiency.

To demonstrate the interpretability of the proposed TM framework, class-wise vote scores and clause activation heatmaps explain the underlying classification decisions. A larger vote score for a particular class indicates stronger evidence supporting the assignment of an input PDF to that class, reflecting greater confidence in the prediction. Furthermore, feature-level contribution analysis is performed to identify the most influential features driving the classification outcome. The performance metrics are defined as follows~\cite{jaiswal2022performance,jaiswal2022non}:
\begin{equation}
\text{Accuracy} = \frac{TP + TN}{TP + TN + FP + FN},
\end{equation}
\begin{equation}
\text{Precision} = \frac{TP}{TP + FP},~~\text{Recall} = \frac{TP}{TP + FN},
\end{equation}
\begin{equation}
\text{F1-score} = \frac{2 \times (\text{Precision} \times \text{Recall})}
{\text{Precision} + \text{Recall}},
\end{equation}
where $TP$, $TN$, $FP$, and $FN$ denote the true positive, true negative, false positive, and false negative, respectively. 

\subsection{Classification Performance} 
\label{clas_ana}
\subsubsection{Data Pre-processing} 
\begin{figure}[t!]
\centering
\includegraphics[width=\columnwidth,height=4.0cm,keepaspectratio]{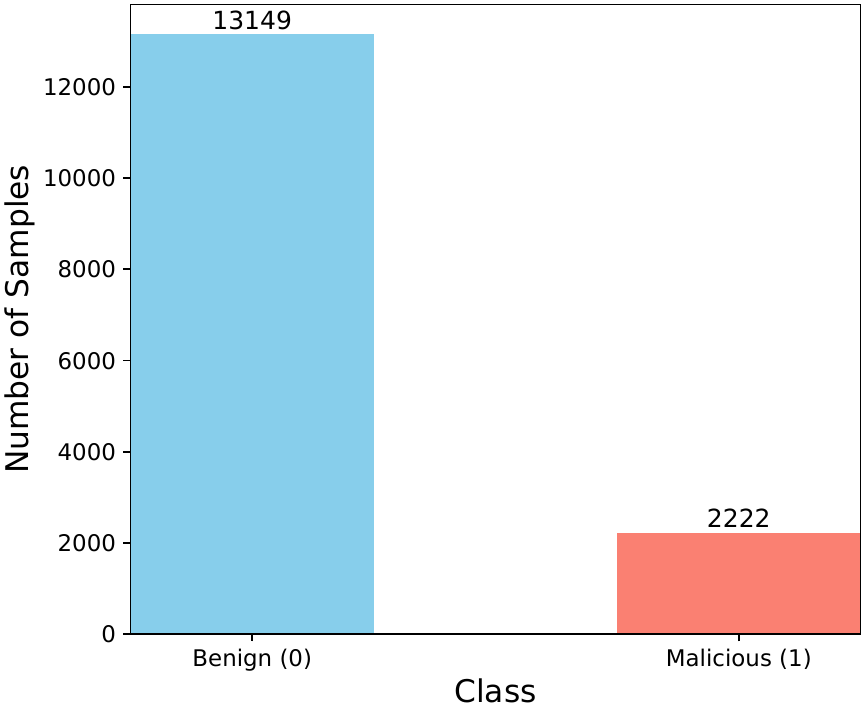} 
\caption{Class imbalance in the dataset.}
\label{clas_imb} \vspace{-2mm}
\end{figure}

\begin{figure}[t!]
\centering
\includegraphics[width=0.9\columnwidth,height=4.0cm,keepaspectratio]{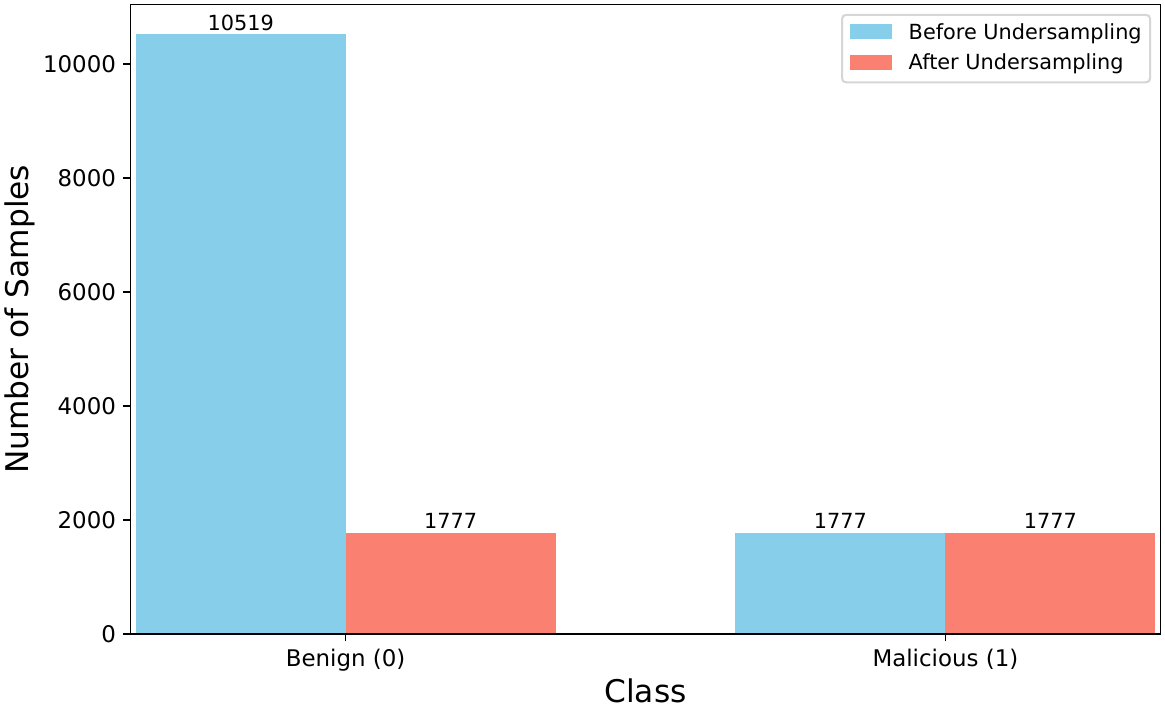} 
\caption{Balanced training classes.}
\label{clas_imb_undersamp} \vspace{-2mm}
\end{figure}

The dataset is imbalanced, as summarized in Table~\ref{summary_sample}. During data preprocessing, 8,966 duplicate samples (36.84\% of the dataset) are identified and removed, resulting in 15,371 unique samples comprising 13,149 benign and 2,222 malicious PDF files. No missing values are observed in the dataset. The class labels are encoded numerically, where 0 denotes benign PDFs, and 1 denotes malicious PDFs. Furthermore, the features \texttt{Acroform}, \texttt{Colors}, and \texttt{BaseEncoding} are excluded because they contain only zero values and therefore do not contribute to the classification process. After preprocessing, the dataset remains highly imbalanced, as illustrated in Fig.~\ref{clas_imb}. Such imbalance can bias the learning process toward the majority class and adversely affect detection performance. To address this issue, the dataset is divided into training (80\%) and testing (20\%) subsets using stratified random sampling with a fixed random seed of 42. Random undersampling is then applied exclusively to the training set to balance the class distribution, as shown in Fig.~\ref{clas_imb_undersamp}, while preserving the original distribution of the test set for unbiased performance evaluation.

\begin{figure}[t!]
\centering
\includegraphics[width=0.9\columnwidth,height=4.0cm,keepaspectratio]{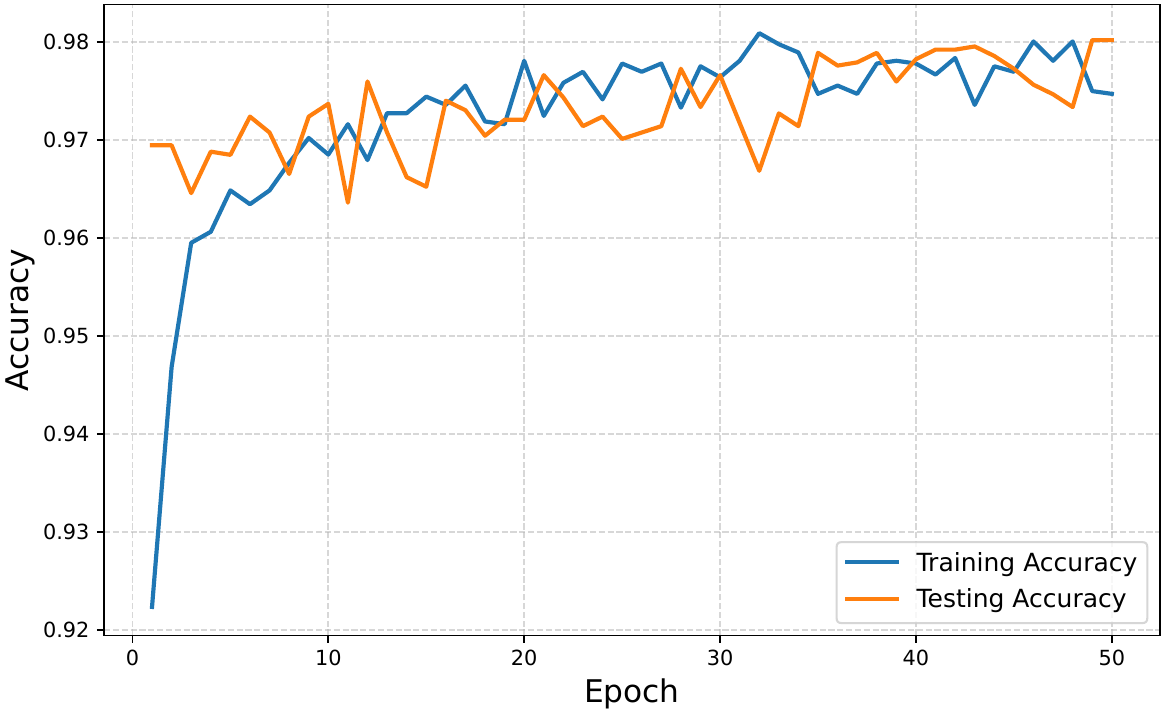} 
\caption{Training and testing accuracy across epochs.}
\label{train_test_acc_TM} \vspace{-3mm}
\end{figure}

\subsubsection{Classifier Training} 
The numerical features are normalized using min-max scaling, where the scaler is fitted on the training set and subsequently applied to the test set to ensure consistent feature scaling and stable model training. Since the TM requires binary-valued inputs for logical rule learning, the normalized features are discretized into intervals using the KBinsDiscretizer~\cite{bhagwat2019applied} and then converted into a binary representation suitable for clause construction. For a fair comparison, the same normalized data, without the binarization step, are used to train the ML classifiers described in Section~\ref{mlc}. All model parameters are selected empirically, and a fixed random seed of 42 is maintained throughout the experiments to ensure reproducibility. The parameter configurations and classification results of the TM and ML models are presented in Tables~\ref{para_TM_ML} and~\ref{mod_perf}, respectively. Additionally, Fig.~\ref{train_test_acc_TM} illustrates the training and testing accuracy of the TM model across the training epochs, demonstrating stable convergence and consistent learning behavior.

\begin{table} [t!]
\caption{Model parameters.}
\label{para_TM_ML} 
\centering
{
\setlength\tabcolsep{4.0pt}
\begin{tabular}{|c|c|}
\hline
Model & Parameters \\
\hline
TM & Binarizer: KBinsDiscretizer, \text{n\_bins}=15, \\ 
& encode=\text{onehot-dense}, strategy=quantile \\
\cline{2-2}
& \text{number\_of\_clauses}=250, $T$=15, $s$=5, \\
& \text{weighted\_clauses}=False, Epochs=50 \\
\hline
DT & criterion=gini \\
\hline
KNN & \text{n\_neighbors=4}, algorithm=brute, \text{leaf\_size}=10 \\
\hline
NB & \text{var\_smoothing}=1e-07 \\
\hline
LR & solver=liblinear, \text{max\_iter}=300 \\
\hline
XGBoost & objective=binary:logistic, \text{eval\_metric}=logloss, \\
& \text{tree\_method}=hist, \text{learning\_rate}=0.2, \\ & \text{max\_depth}=10, \text{n\_estimators=100} \\
\hline
LGBM & objective=binary, \text{learning\_rate}=0.2, \text{n\_estimators=200}, \\ 
& \text{num\_leaves}=20 \\
\hline
\end{tabular} \vspace{-1mm}
}
\end{table}

\begin{table} [t!]
\caption{Model performance.}
\label{mod_perf} 
\centering
{
\setlength\tabcolsep{3.0pt}
\begin{tabular}{|c|c|c|c|c|c|}
\hline
Model & Accuracy & Precision & Recall & \text{F1-score} & Inference time \\
& (in \%) & (in \%) & (in \%) & (in \%) & (in $\mu$s) \\
\hline
TM & 98.02 & 96.03 & 95.95 & 95.99 & 2.853 \\
\hline
DT & 94.41 & 86.45 & 94.77 & 89.89 & 0.044 \\
\hline
KNN & 94.31 & 86.77 & 92.75 & 89.39 & 3.659 \\
\hline
NB & 86.57 & 73.92 & 59.76 & 62.37 & 0.115 \\
\hline
LR & 79.74 & 66.06 & 74.43 & 68.13 & 0.032 \\
\hline
XGBoost & 97.37 & 93.06 & 97.06 & 94.92 & 0.319 \\
\hline
LGBM & 97.30 & 92.88 & 97.02 & 94.80 & 15.274 \\
\hline
\end{tabular} \vspace{-1mm}
}
\end{table}

\begin{table} [t!]
\caption{TM performance by class.}
\label{class_wise_perf_TM} 
\centering
{
\setlength\tabcolsep{5.0pt}
\begin{tabular}{|c|c|c|c|}
\hline
Class & Precision (in \%) & Recall (in \%) & \text{F1-score} (in \%) \\
\hline
Benign & 98.82 & 98.86 & 98.84 \\
\hline
Malicious & 93.24 & 93.03 & 93.14 \\
\hline
\end{tabular} 
}
\end{table}

\begin{figure}[t!]
\centering
\includegraphics[width=\columnwidth,height=5.0cm,keepaspectratio]{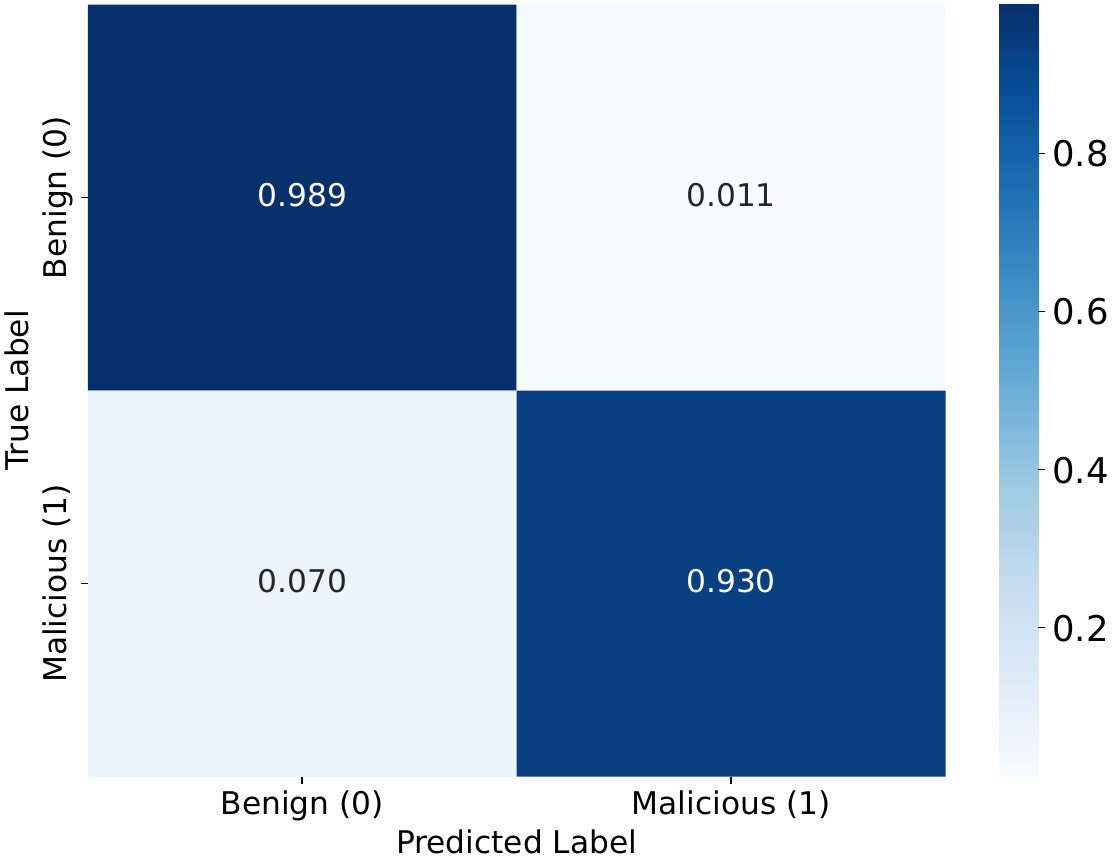} 
\caption{TM confusion matrix.}
\label{cm_TM} \vspace{-3mm}
\end{figure}

Table~\ref{mod_perf} shows that the proposed TM model achieves the highest accuracy of 98.02\%, along with the best precision (96.03\%) and F1-score (95.99\%), demonstrating balanced classification performance across both benign and malicious PDF classes. XGBoost and LightGBM achieve competitive accuracies of 97.37\% and 97.30\%, respectively, but achieve slightly lower performance than the proposed TM model. Although Logistic Regression requires the lowest inference time of 0.032~$\mu$s per sample, its accuracy is considerably lower (79.74\%). In contrast, the proposed TM model requires 2.853~$\mu$s per sample, making it faster than both LightGBM and KNN while maintaining the highest detection performance. Overall, the proposed TM model offers an effective balance between classification accuracy and computational efficiency, making it suitable for real-time PDF malware detection.

Table~\ref{class_wise_perf_TM} shows that the high evaluation metrics achieved for both benign and malicious PDF classes demonstrate the effectiveness of the TM model in accurately detecting PDF malware while maintaining balanced classification performance.

Next, Fig.~\ref{cm_TM} presents the confusion matrix of the proposed TM model. The model correctly classifies 98.9\% of benign PDF documents and 93.0\% of malicious PDF documents. This corresponds to a false positive rate (FPR) of only 1.1\%, indicating that very few benign PDFs are incorrectly misclassified as malicious, and a false negative rate (FNR) of 7.0\%, showing that only a small proportion of malicious PDFs are misclassified as benign. These results demonstrate the strong classification capability of the TM model, with high detection accuracy and low misclassification rates for both classes.

\subsubsection{TM Interpretability} 
\begin{figure}[t!]
\centering
\includegraphics[width=0.95\columnwidth,height=4.0cm,keepaspectratio]{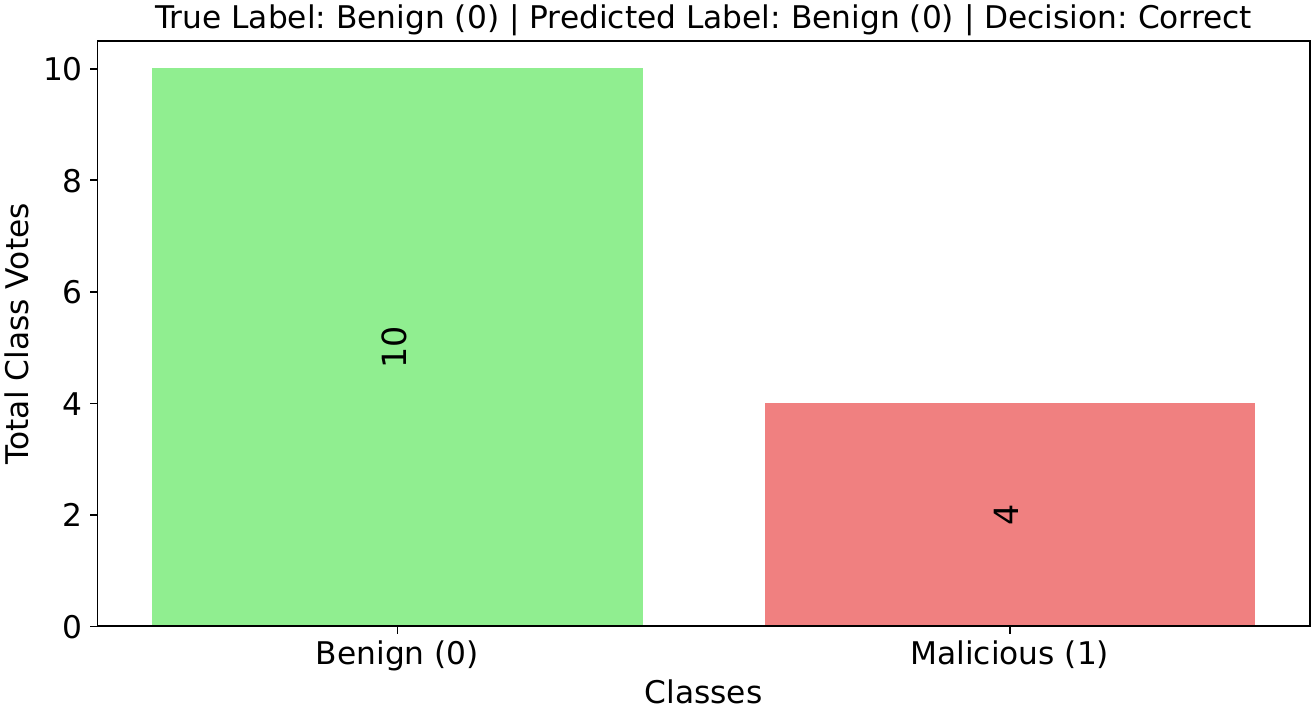} 
\caption{Class-wise votes of a Benign sample.}
\label{cls_vote_benign} \vspace{-2mm}
\end{figure}

\begin{figure}[t]
\centering
\includegraphics[width=0.95\columnwidth,height=4.0cm,keepaspectratio]{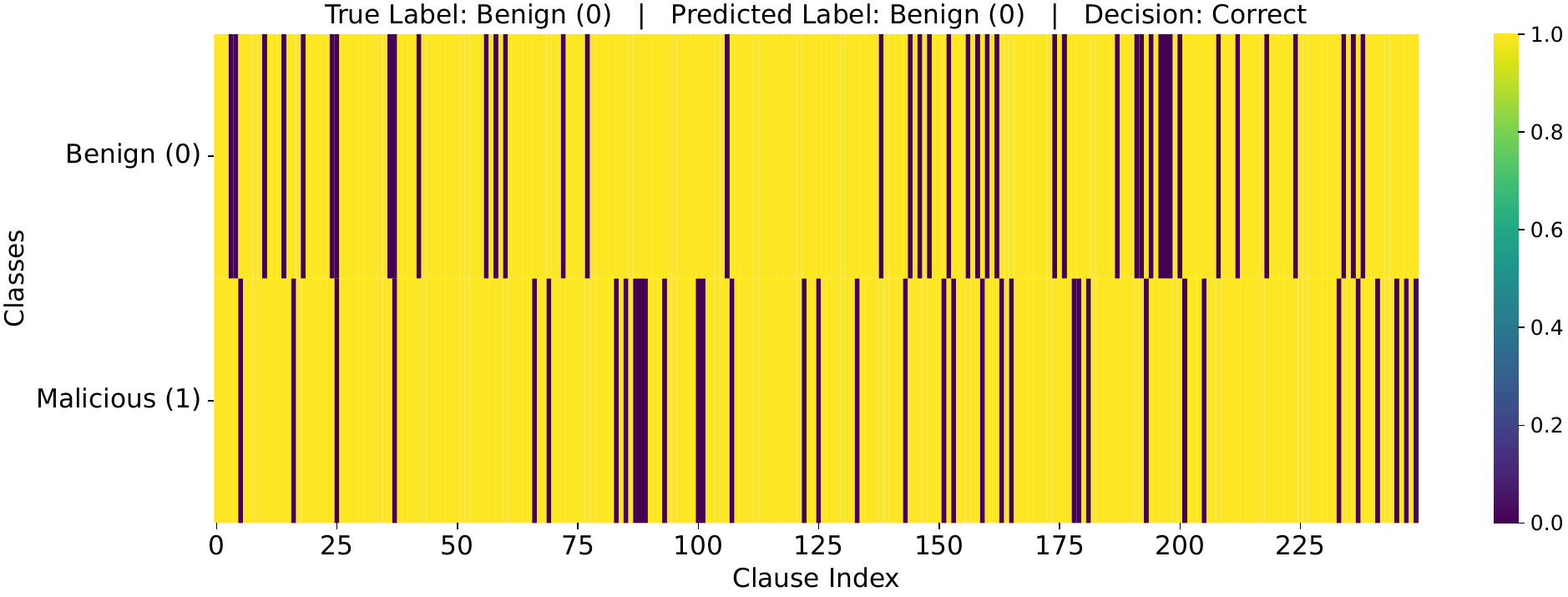} 
\caption{Clause activation heatmap of the same Benign sample.}
\label{cls_heatmap_benign} \vspace{-2mm}
\end{figure}

\begin{figure}[t!]
\centering
\includegraphics[width=0.95\columnwidth,height=4.0cm,keepaspectratio]{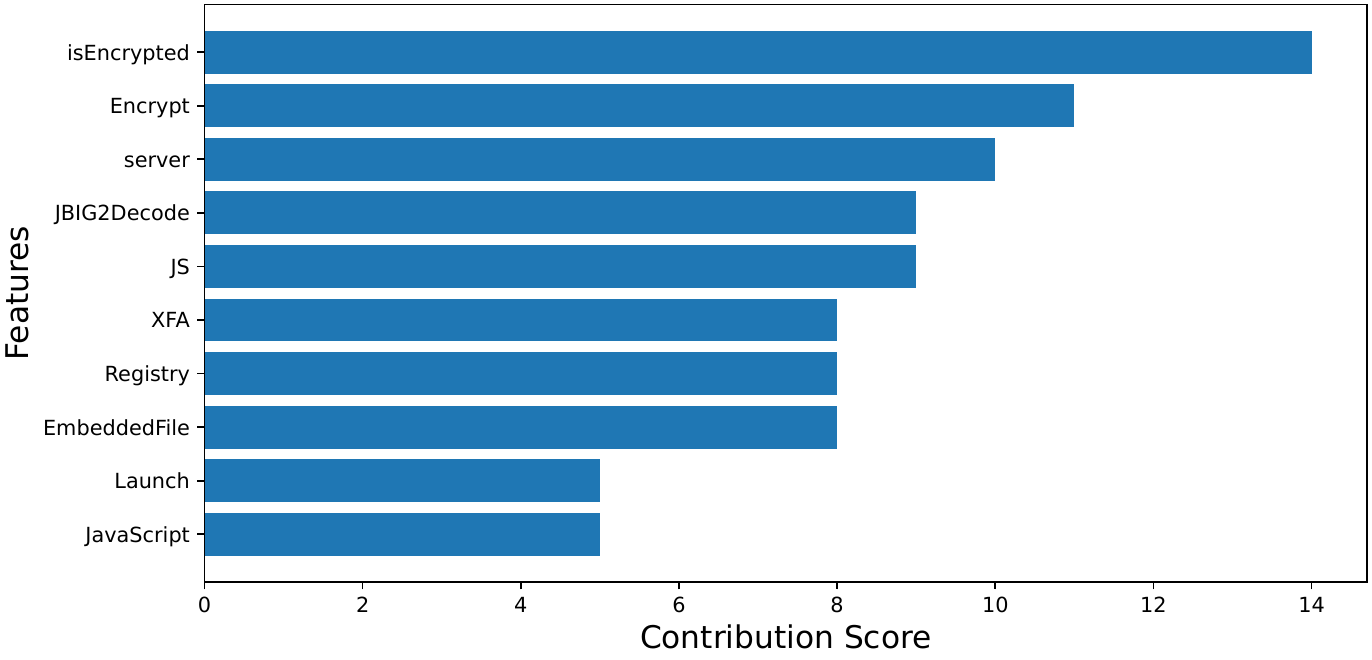} 
\caption{Top ten features of the same Benign sample.}
\label{top10_feat_benign} \vspace{-3mm}
\end{figure}

\begin{figure*}[t!]
\centering
\includegraphics[width=\textwidth,height=4.75cm,keepaspectratio]{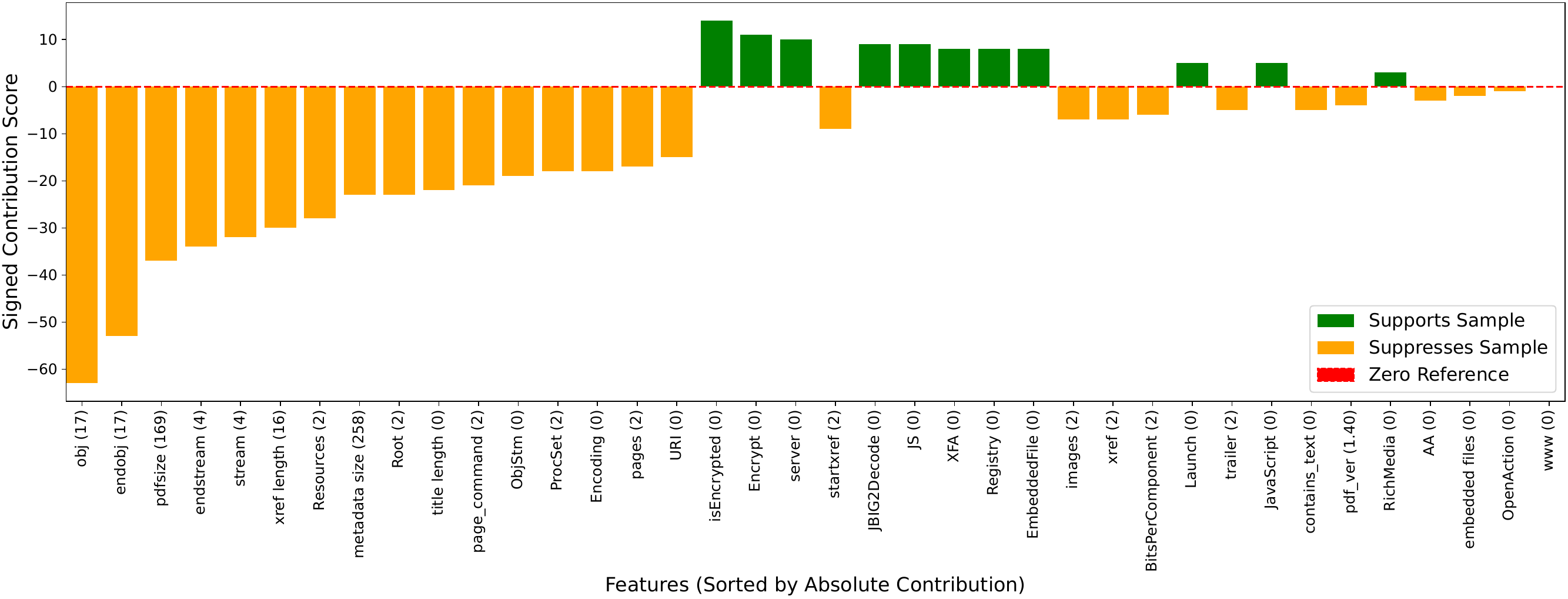}
\caption{Feature-level contribution of the same Benign sample as in Fig.~\ref{cls_vote_benign}. Feature values are shown in parentheses.}
\label{feat_contri_benign} \vspace{-2mm}
\end{figure*}

To illustrate the interpretability of the proposed TM framework, Figs.~\ref{cls_vote_benign}, \ref{cls_heatmap_benign}, \ref{top10_feat_benign}, and \ref{feat_contri_benign} present the class-wise vote scores, clause activation heatmap, top ten contributing features, and feature-level contributions for a benign test sample, respectively.

Figure~\ref{cls_vote_benign} shows that the benign class receives the highest class vote (10), resulting in the correct classification of the input PDF as benign. As illustrated in Fig.~\ref{cls_heatmap_benign}, each cell represents the activation state of a clause for the selected test sample, where yellow (1) denotes an active clause and dark purple (0) indicates an inactive clause. The benign class exhibits a larger number of activated clauses than the malicious class, leading to a higher accumulated class vote and, consequently, the correct prediction. Figure~\ref{top10_feat_benign} presents the ten most influential features contributing to the classification, with \texttt{isEncrypted} emerging as the most significant feature for the selected benign PDF sample. Finally, Fig.~\ref{feat_contri_benign} illustrates the feature-level contributions, where positive contribution values (green) support the benign sample prediction, whereas negative contribution values (orange) suppress evidence for the benign class. The corresponding feature values are shown in parentheses to provide additional context for the explanation. These results demonstrate the transparent and interpretable decision-making capability of the TM model. 

\begin{figure*}[t!]
\centering
\includegraphics[width=\textwidth,height=4.75cm,keepaspectratio]{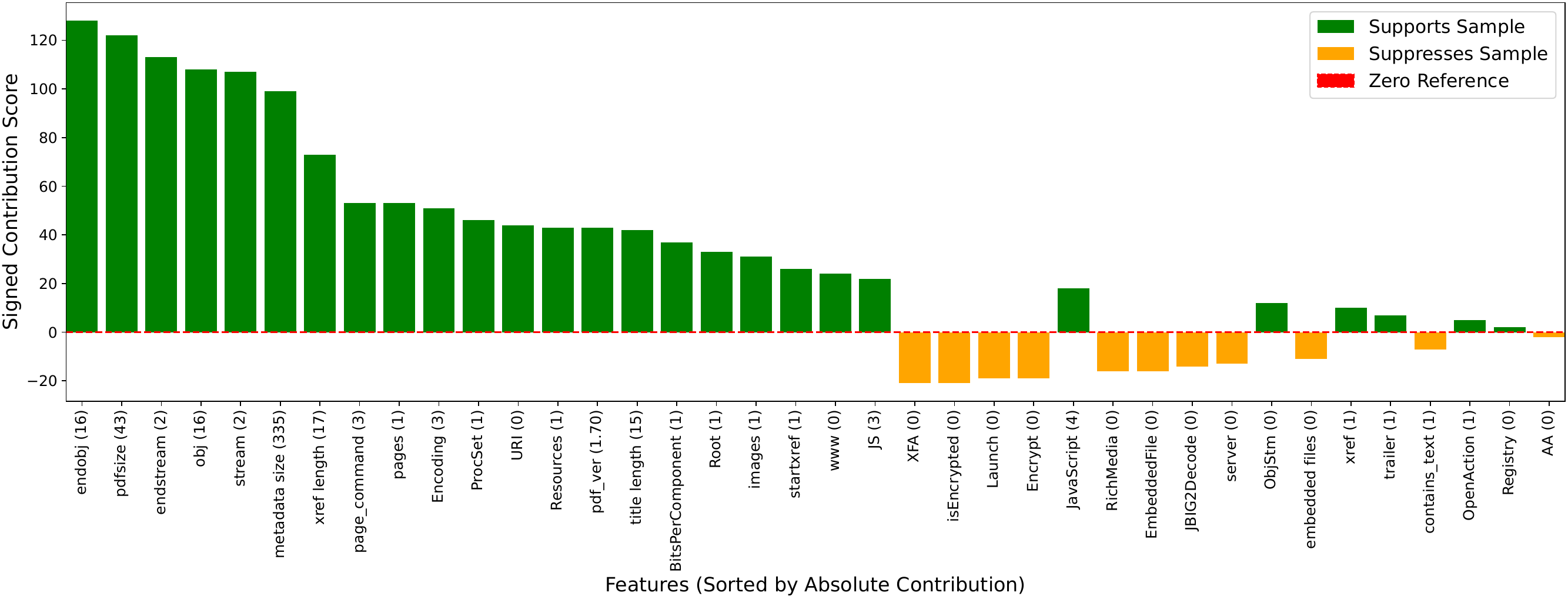} 
\caption{Feature-level contribution of the same Malicious sample as in Fig.~\ref{cls_vote_mal_samp}. Feature values are shown in parentheses.}
\label{feat_contri_mal122} \vspace{-3mm}
\end{figure*}

\begin{figure}[t!]
\centering
\includegraphics[width=0.95\columnwidth,height=3.75cm,keepaspectratio]{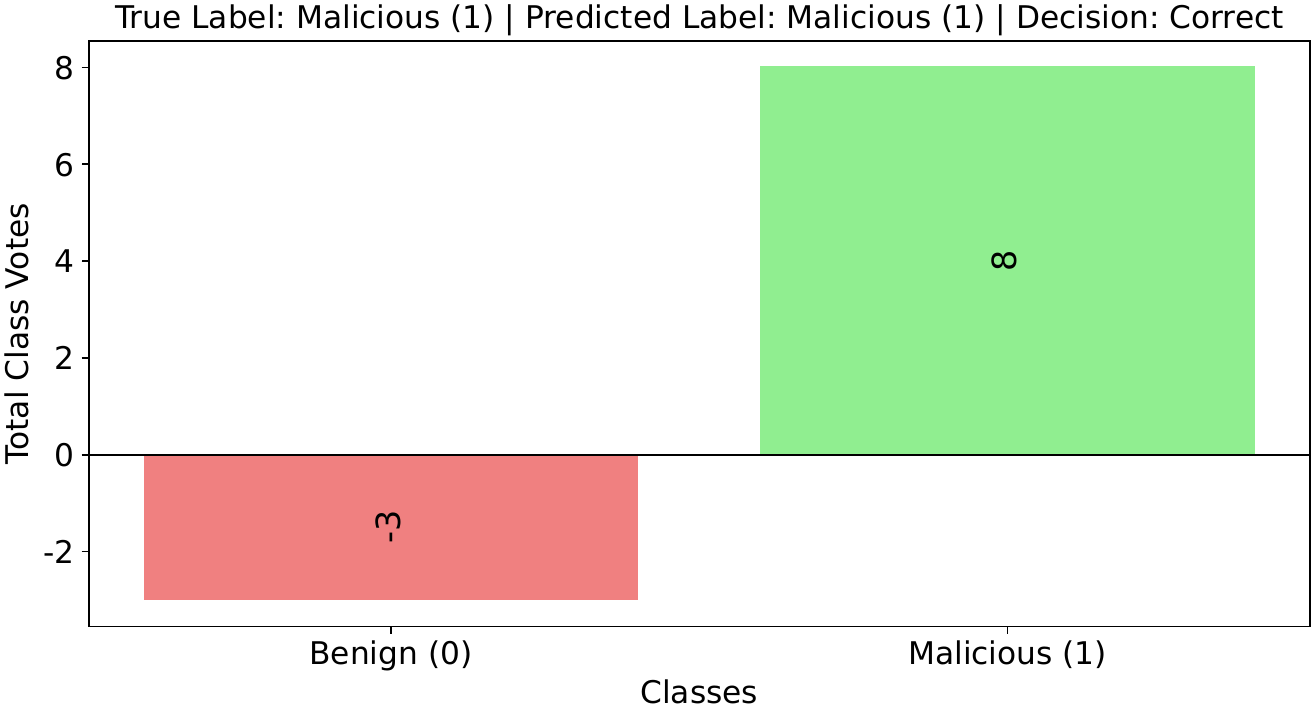} 
\caption{Class-wise votes of a Malicious sample.}
\label{cls_vote_mal_samp} \vspace{-3mm}
\end{figure}

\begin{figure}[t!]
\centering
\includegraphics[width=0.95\columnwidth,height=3.75cm,keepaspectratio]{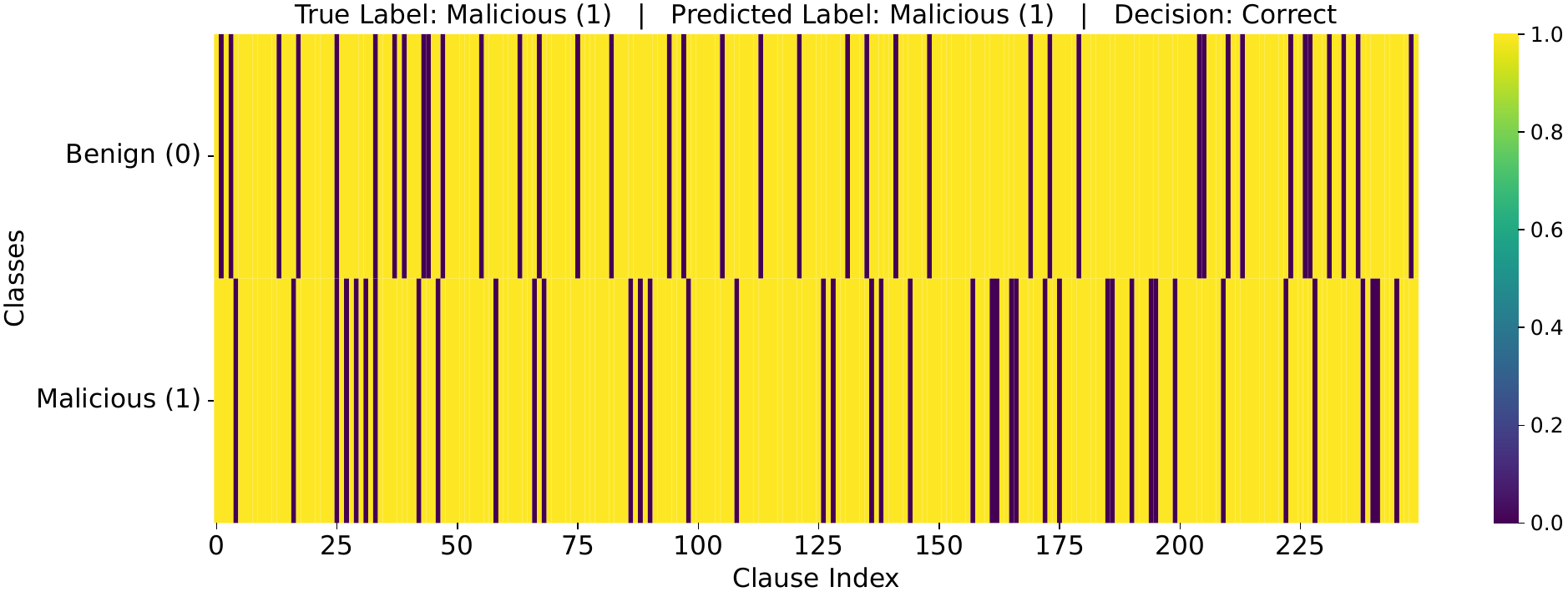} 
\caption{Clause activation heatmap of same Malicious sample.}
\label{cls_heatmap_mal_samp} \vspace{-3mm}
\end{figure}

Similarly, Fig.~\ref{cls_vote_mal_samp} shows that the malicious class receives the highest class vote (8), resulting in the correct classification of the input PDF as malicious. Figure~\ref{cls_heatmap_mal_samp} also shows that the malicious class has more activated clauses than the benign class, yielding a higher cumulative vote and correct prediction. This observation is further supported by the feature-level contributions shown in Fig.~\ref{feat_contri_mal122}, highlighting the transparent and interpretable decision-making capability of the TM model.

\begin{figure}[t!]
\centering
\includegraphics[width=0.95\columnwidth,height=3.75cm,keepaspectratio]{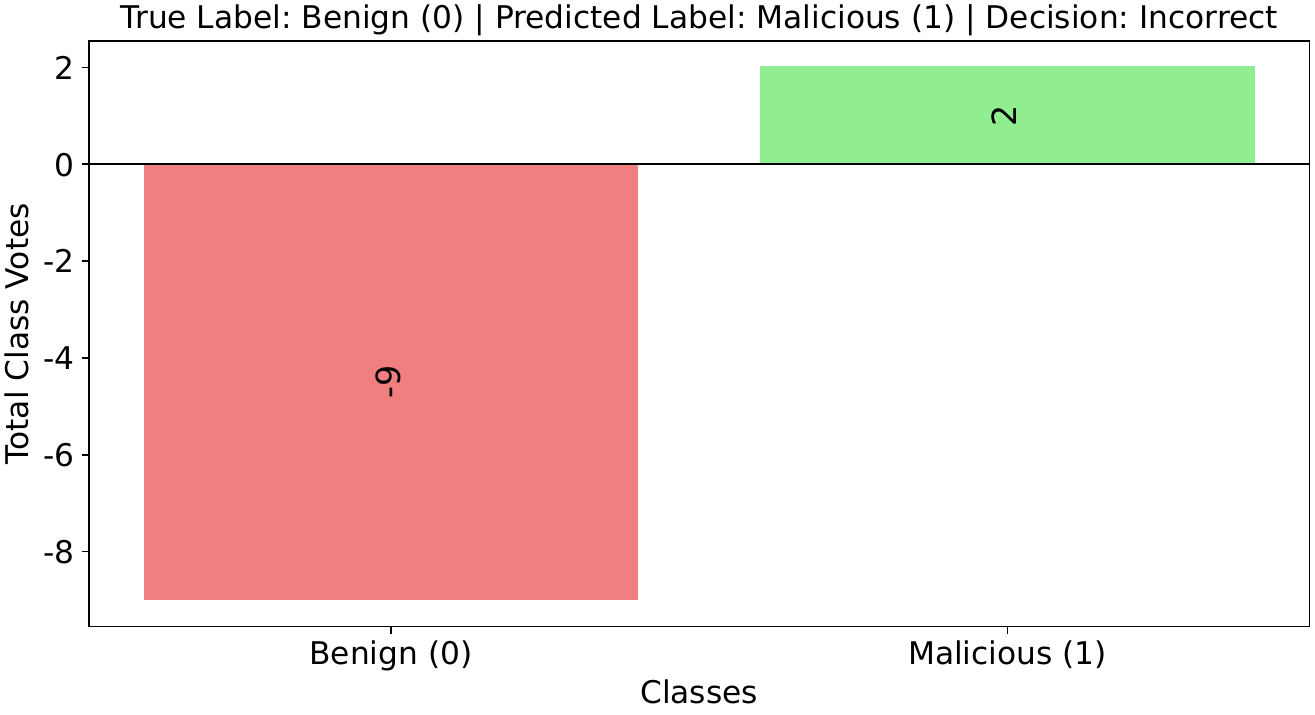} 
\caption{Class-wise votes of another Benign sample.}
\label{cls_vote_ben_samp521} \vspace{-2mm}
\end{figure}

\begin{figure}[t!]
\centering
\includegraphics[width=0.95\columnwidth,height=3.75cm,keepaspectratio]{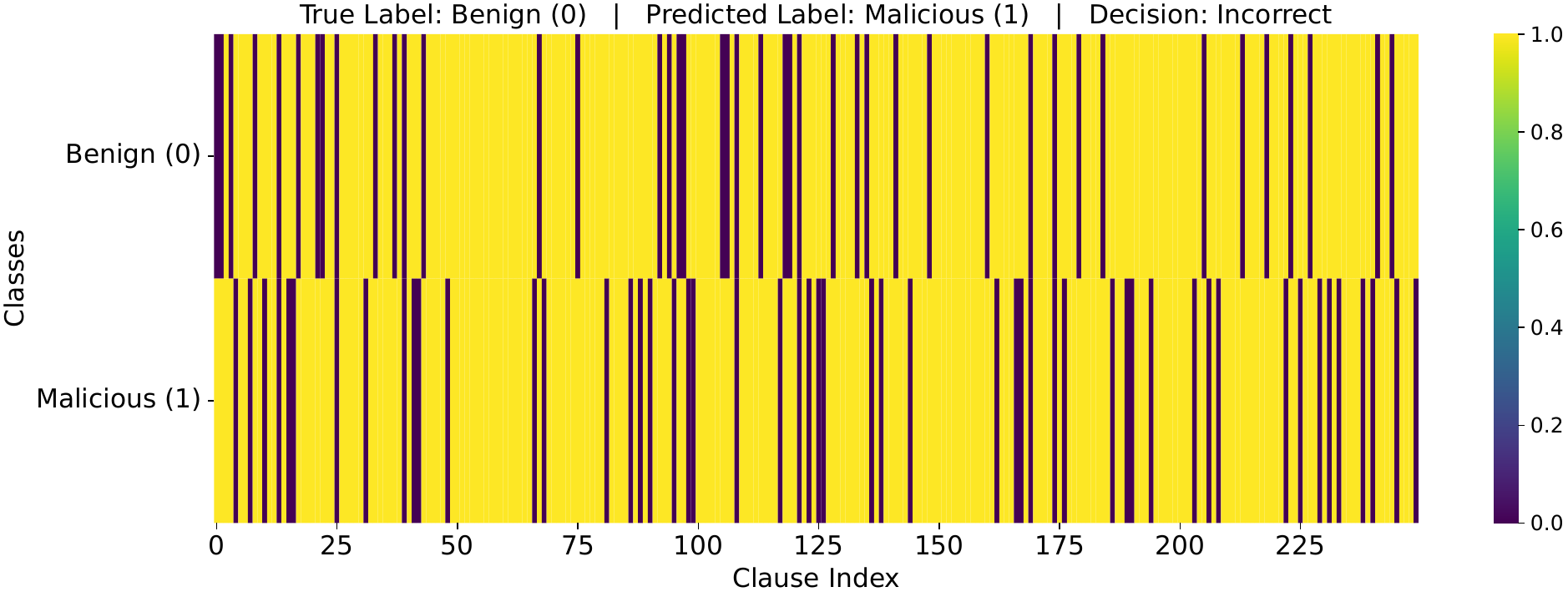} 
\caption{Clause activation heatmap of the sample as in Fig.~\ref{cls_vote_ben_samp521}.}
\label{cls_heatmap_ben_samp521} \vspace{-3mm}
\end{figure}

\begin{figure*}[t!]
\centering
\includegraphics[width=\textwidth,height=4.75cm,keepaspectratio]{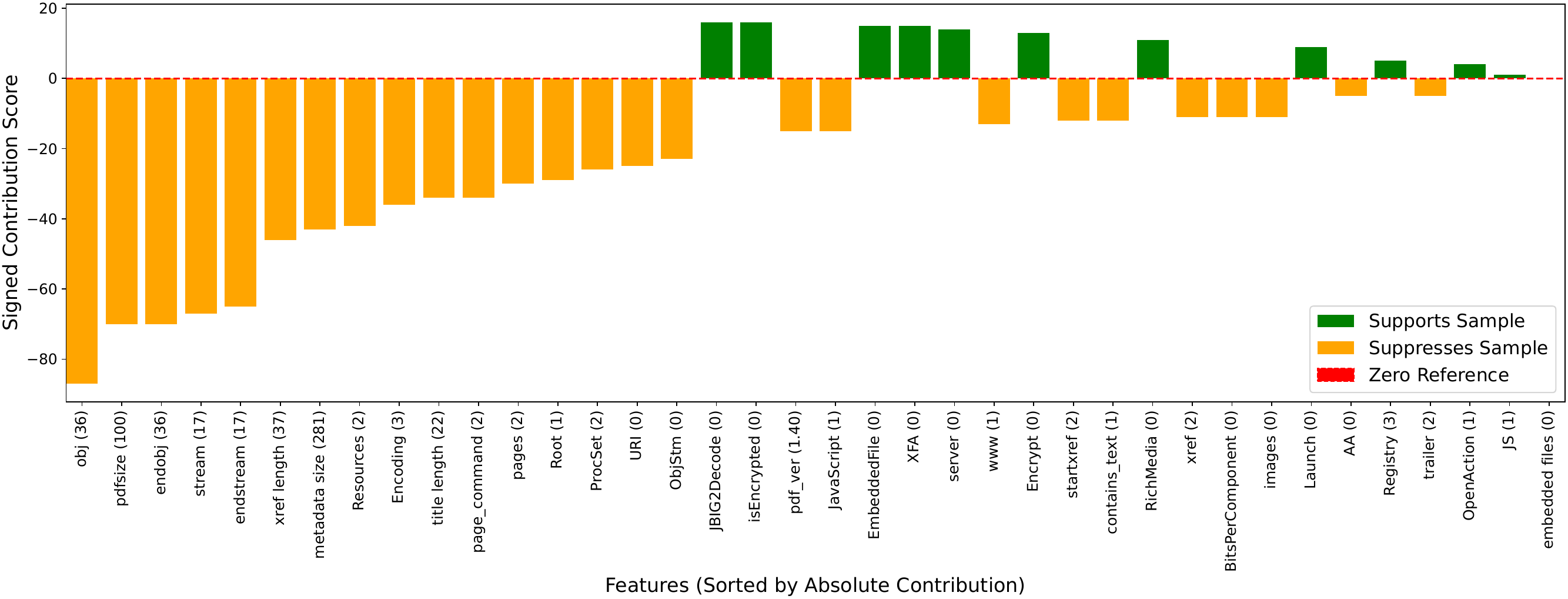} 
\caption{Feature-level contribution of another benign sample as in Fig.~\ref{cls_vote_ben_samp521}. Feature values are shown in parentheses.}
\label{feat_contri_mal521} 
\end{figure*}

In contrast, Fig.~\ref{cls_vote_ben_samp521} shows a misclassification case in which a benign sample is incorrectly classified as malicious. The malicious class attains the highest class vote (2), whereas the correct benign class receives a vote of -9. This misclassification is caused by overlapping feature characteristics between benign and malicious samples, leading the TM model to assign a higher confidence to the malicious class. This behavior is further supported by the clause activation heatmap shown in Fig.~\ref{cls_heatmap_ben_samp521}, where the malicious class exhibits a larger number of activated clauses than the benign class. Similarly, Fig.~\ref{feat_contri_mal521} illustrates the suppressing feature contributions. 


\subsection{Edge Device Evaluation}
\label{edge_ana}
Table~\ref{edge_perf} demonstrates that the proposed TM model achieves efficient on-device inference with a latency of 10.849~$\mu$s while maintaining moderate CPU utilization (30.8\%) and a compact model size (154.04~KB) compared with XGBoost (249.12~KB), LightGBM (468.18~KB), and KNN (556.11~KB). Although DT, LR, and NB exhibit lower inference times, they achieve considerably lower detection performance than the TM model. These results demonstrate the suitability of the proposed TM framework for real-time PDF malware detection on resource-constrained edge devices.

\begin{table} [t!]
\caption{Edge performance.}
\label{edge_perf} 
\centering
{
\setlength\tabcolsep{4.0pt}
\begin{tabular}{|c|c|c|c|c|}
\hline
Model & Inference time & Memory usage & CPU usage & Model size \\
& (in $\mu$s) & (in KB) & (in \%) & (in KB) \\
\hline
TM & 10.849 & 1872 & 30.8 & 154.04 \\
\hline
DT & 0.508 & 112 & 0.1 & 24.05 \\
\hline
KNN & 33.466 & 1872 & 69.2 & 556.11 \\
\hline
NB & 3.227 & 1120 & 0.2 & 1.94 \\
\hline
LR & 0.812 & 192 & 66.7 & 1.10 \\
\hline
XGBoost & 5.664 & 928 & 16.7 & 249.12 \\
\hline
LGBM & 7.814 & 528 & 44.4 & 468.18 \\
\hline
\end{tabular} \vspace{-3mm}
}
\end{table}

\subsection{State-of-the-Art Comparison} 
\label{comp_sota}
Table~\ref{comp_table} shows that, although the compared methods are evaluated on different datasets, the proposed TM framework achieves competitive accuracy while providing intrinsic interpretability and low inference time. Moreover, deployment on a Raspberry Pi shows its capability for real-time, on-device PDF malware detection. Consequently, the proposed framework enables a balanced trade-off among detection performance, computational efficiency, transparency, and edge deployability, making it a practical solution for PDF malware detection.

\begin{table*} [t!]
\caption{Comparison with related works.}
\label{comp_table} 
\centering
{
\setlength\tabcolsep{5.0pt}
\begin{tabular}{|c|c|c|c|c|c|c|}
\hline
Reference & Dataset & Method & Accuracy & Inference Time & Interpretable & Edge Deployment \\
\hline
Paper~\cite{liu2025vapd} & Contagio & VAPD & 99.54\% & - & No & No \\
\hline
Paper~\cite{abu2022pdf} & Evasive-2022 & Optimized DT & 98.84\% & 2.174 $\mu$s & No & No \\
\hline
Paper~\cite{chbib2024leveraging} & Contagio & ML Ensemble & 93.00\% & - & Yes & No \\
\hline
Proposed & RIT-PDFMal-2026 & TM & 98.02\% & 2.853 $\mu$s & Yes & Yes \\
\hline
\end{tabular} \vspace{-3mm}
}
\end{table*}

\section{Conclusions and Future Work} 
\label{con_fut}
This paper presents an interpretable Tsetlin Machine (TM)-based framework for PDF malware detection. The proposed TM framework extracts salient features directly from PDF documents through static analysis without executing the files and employs rule-based learning to accurately classify benign and malicious PDF files. Experimental results on the RIT-PDFMal-2026 dataset demonstrate that the proposed TM framework outperforms several conventional ML classifiers while providing intrinsic interpretability of its classification decisions. In addition, deployment of the TM model on a Raspberry Pi demonstrates the practicality of real-time on-device PDF malware detection in resource-constrained environments. This enhances the transparency and trustworthiness of the classification process, making the proposed TM framework a promising solution for practical PDF malware detection. Nevertheless, the study is limited to a single public dataset, and the interpretability analysis is demonstrated using representative case studies rather than a formal user-based evaluation. Future work will focus on extending the evaluation to diverse PDF malware datasets encompassing a wider range of malware families and validating its interpretability through user studies.

\section*{Acknowledgement}
This publication has emanated from the research project SecureIoTM: Ultra-low-energy IoT Intrusion Detection Systems using Logic-based Tsetlin Machines, under Grant Number 342167, funded by the Research Council of Norway.

\bibliographystyle{IEEEtran}
\bibliography{references}

@misc{report_cloudfiles,
  author = {CloudFiles},
  title = {{Digital files presence in the world, 2025}},
  year = {Accessed on July 05, 2026},
  url = {https://www.cloudfiles.io/blog/how-many-files-are-there-in-the-world}
}

@article{singh2020malware,
  title={{Malware Detection in PDF and Office Documents: A Survey}},
  author={Singh, Priyansh and Tapaswi, Shashikala and Gupta, Sanchit},
  journal={{Information Security Journal: A Global Perspective}},
  volume={29},
  number={3},
  pages={134--153},
  year={2020},
  publisher={Taylor \& Francis}
}

@misc{report_reis,
  author = {Reis Informatica},
  title = {{Malicious PDF Attacks on Microsoft Windows, 2026}},
  year = {Accessed on July 05, 2026},
  url = {https://reisinformatica.com/malicious-pdf-attacks-on-microsoft-windows-2026-protection-guide-for-canadian-businesses/}
}

@inproceedings{liu2025vapd,
  title={{VAPD: An Anomaly Detection Model for PDF Malware Forensics with Adversarial Robustness}},
  author={Liu, Side and Ming, Jiang and Zhou, Yilin and Fu, Jianming and Peng, Guojun},
  booktitle={34th USENIX Security Symposium},
  pages={4759--4778},
  year={2025}
}

@article{abu2022pdf,
  title={{PDF Malware Detection Based on Optimizable Decision Trees}},
  author={Abu Al-Haija, Qasem and Qattous, Hazem},
  journal={Electronics},
  volume={11},
  number={19},
  pages={3142},
  year={2022},
  publisher={MDPI}
}

@inproceedings{chbib2024leveraging,
  title={{Leveraging Machine Learning-Based PDF Malware Detection in Snort}},
  author={Chbib, Fadlallah and Khatoun, Rida},
  booktitle={Int. Conference on Electrical, Computer, Communications and Mechatronics Engineering},
  pages={1--6},
  year={2024},
  organization={IEEE}
}

@inproceedings{rong2020transnet,
  title={{TransNet: Unseen Malware Variants Detection using Deep Transfer Learning}},
  author={Rong, Candong and Gou, Gaopeng and Cui, Mingxin and Li, Zhen and Guo, Li},
  booktitle={Int. Conf. on Security and Privacy in Communication Systems},
  pages={84--101},
  year={2020},
  organization={}
}

@inproceedings{lundberg2017unified,
  title={{A Unified Approach to Interpreting Model Predictions}},
  author={Lundberg, Scott M and Lee, Su-In},
  booktitle={NIPS},
  pages={1--10},
  year={2017}
}

@article{granmo2018tsetlin,
  title={{The Tsetlin Machine--A Game Theoretic Bandit Driven Approach to Optimal Pattern Recognition with Propositional Logic}},
  author={Granmo, Ole-Christoffer},
  journal={arXiv preprint arXiv:1804.01508},
  pages={1--42},
  year={2018}
}

@article{kundu2026comprehensive,
  title={{A Comprehensive Review of Tsetlin Machines: Concepts, Applications, Analysis, and the Future}},
  author={Kundu, Souraja and Mishra, Saras Mani and Trivedi, Gaurav and Merchant, Farhad},
  journal={IEEE IoT Journal},
  volume={13},
  number={10},
  pages={20105--20127},
  year={2026},
  publisher={IEEE}
}

@article{jaiswal2025leveraging,
  title={{Leveraging Transfer learning for Radio Map Estimation via Mixture of Experts}},
  author={Jaiswal, Rahul Kumar and Elnourani, Mohamed and Deshmukh, Siddharth and Beferull-Lozano, Baltasar},
  journal={IEEE TCCN},
  volume={12},
  pages={846--863},
  year={2025},
  publisher={IEEE}
}

@inproceedings{jaiswal2023location,
  title={{Location-free Indoor Radio Map Estimation using Transfer learning}},
  author={Jaiswal, Rahul and Elnourani, Mohamed and Deshmukh, Siddharth and Beferull-Lozano, Baltasar},
  booktitle={97th Vehicular Technology Conference},
  pages={1--7},
  year={2023},
  organization={IEEE}
}

@article{jaiswal2025data,
  title={{A Data-driven Transfer Learning Method for Indoor Radio Map Estimation}},
  author={Jaiswal, Rahul Kumar and Elnourani, Mohamed and Deshmukh, Siddharth and Beferull-Lozano, Baltasar},
  journal={IEEE TVT},
  volume={75},
  number={3},
  pages={4261--4277},
  year={2026},
  publisher={IEEE}
}

@inproceedings{chen2016xgboost,
  title={{Xgboost: A Scalable Tree Boosting System}},
  author={Chen, Tianqi and Guestrin, Carlos},
  booktitle={22nd ACM SIGKDD International Conference on Knowledge Discovery and Data Mining},
  pages={785--794},
  year={2016}
}

@inproceedings{ke2017lightgbm,
  title={{Lightgbm: A Highly Efficient Gradient Boosting Decision Tree}},
  author={Ke, Guolin and Meng, Qi and Finley, Thomas},
  booktitle={NIPS},
  pages={1--9},
  year={2017}
}

@book{alpaydin2020introduction,
  title={{Introduction to Machine Learning}},
  author={Alpaydin, Ethem},
  year={2020},
  publisher={MIT press}
}

@book{breiman2017classification,
  title={{Classification and Regression Trees}},
  author={Breiman, Leo and Friedman, Jerome and Olshen, Richard A and Stone, Charles J},
  year={2017},
  publisher={Chapman and Hall/CRC}
}

@book{hosmer2013applied,
  title={Applied Logistic Regression},
  author={Hosmer Jr, David W and Lemeshow, Stanley and Sturdivant, Rodney X},
  year={2013},
  publisher={John Wiley \& Sons}
}

@inproceedings{jaiswal2022performance,
  title={{Performance Analysis of Voice Activity Detector in Presence of Non-stationary Noise}},
  author={Jaiswal, Rahul},
  booktitle={11th International Conf. on Robotics, Vision, Signal Processing and Power Applications},
  pages={59--65},
  year={2022},
  organization={}
}

@book{bhagwat2019applied,
  title={{Applied Deep Learning with Keras}},
  author={Bhagwat, Ritesh and Abdolahnejad, Mahla and Moocarme, Matthew},
  year={2019},
  publisher={{Packt Publishing Ltd}}
}

@article{mishra2017handling,
  title={{Handling Imbalanced Data: SMOTE vs Random Undersampling}},
  author={Mishra, Satwik},
  journal={International Research Journal of Engineering and Technology},
  volume={4},
  number={8},
  pages={317--320},
  year={2017}
}

@article{alani2026rit,
  title={{RIT-PDFMal-2026: A Comprehensive Benchmark Dataset for PDF Malware Detection}},
  author={Alani, Mohammed M and Damiani, Ernesto},
  journal={IEEE Access},
  volume={14},
  pages={97841--97855},
  year={2026},
  publisher={IEEE}
}

@misc{link_vt,
  author = {VirusTotal},
  title = {{Malicious Sample}},
  year = {Accessed on July 05, 2026},
  url = {https://www.virustotal.com/gui/home/upload}
}

@article{jaiswal2022non,
  title={{Non-intrusive Speech Quality Assessment using Context-aware Neural Networks}},
  author={Jaiswal, Rahul Kumar and Dubey, Rajesh Kumar},
  journal={International Journal of Speech Technology},
  volume={25},
  number={4},
  pages={947--965},
  year={2022},
  publisher={Springer}
}

@article{mathe2024comprehensive,
  title={{A Comprehensive Review on Applications of Raspberry Pi}},
  author={Mathe, Sudha Ellison and Kondaveeti, Hari Kishan and Vappangi, Suseela and Vanambathina, Sunny Dayal and Kumaravelu, Nandeesh Kumar},
  journal={Computer Science Review},
  volume={52},
  pages={100636},
  year={2024},
  publisher={Elsevier}
}

\end{document}